# Theorizing information sources for hope: belief, desire, imagination, and metacognition

Tim Gorichanaz
College of Computing & Informatics
Drexel University
Philadelphia, Pennsylvania, USA

## Abstract

***Introduction.*** Hope is a positive attitude oriented toward a possible (yet uncertain), desired outcome. Though hope is a virtue, hopelessness is widespread and seems related not only to current events but also to information about current events. This paper examines how hope can be sparked through information.
***Method.*** This study uses the philosophical methods of conceptual analysis and design to advance a theoretical argument.
***Analysis.*** First, a conceptualization of hope is offered, drawing on work primarily in virtue ethics. Then, four types of information sources for hope are theorized, building on and synthesizing work from philosophy and psychology.
***Results.*** Four categories of information source conducive to hopefulness are identified: information for forming beliefs about the past or future; information for engaging the moral imagination regarding possibilities for the future; information for sparking desire for particular moral outcomes; and information for metacognition, or about how we become informed with respect to hope.
***Conclusions.*** Hope is, in many cases, responsive to information. This suggests a moral opportunity for information professionals and scholars to work toward connecting people with information for hope, particularly in difficult times. Avenues for further research, particularly in information behavior and practices, are suggested.
***Topic areas***: History and philosophy of information, Information behavior and practices
***Keywords***: information ethics, hope, information sources

*Cease to lament for that thou canst not help,*
*And study help for that which thou lament'st…*
*Hope is a lover's staff; walk hence with that*
*And manage it against despairing thoughts.*

—William Shakespeare, *The Two Gentlemen of Verona*, III.1

## Introduction

Perhaps it is always the case in life, but these days it seems easy to feel hopeless. In early 2020, the coronavirus pandemic erupted from a global pot that was already simmering with sociopolitical unrest. There have been ups since then, but apparently more downs. As I was writing the first draft of this paper in late 2021, the omicron variant of the coronavirus had just been discovered, raising countless fearful questions. Months later, as I am working on revisions, Vladimir Putin's forces have invaded Ukraine and a bloody war has been ongoing for over a month. No doubt by the time you read these words, some new cause of concern is on your mind. There are many sources for hopelessness besides the pandemic and war: employment and economic challenges, political and ideological battles, harrowing news cycles, the looming threat of climate change… It is no wonder that hopelessness seems to be on the rise.

And this is not just my own impression. According to two reports from spring 2021, for example, 18 percent of U.K. adults reported feeling hopeless about the pandemic (Mental Health Foundation, 2021) and about half of U.S. young adults ages 18–29 reported feeling '*down, depressed or hopeless*' (Harvard Kennedy School Institute of Politics, 2021). Besides hopelessness, we are likewise bearing witness to troubling numbers of people reporting loneliness, anxiety, depression and suicidal ideation (Vahratian et al., 2021).

While hopelessness may seem a reasonable response to certain difficulties, it may actually make our difficulties worse. In many situations, there are actions we can take, however small, that will improve things. Other situations may require concerted collective action, yet amelioration is still a possibility. But if we are fatalistic about a certain outcome, that is, if we are hopeless, then by definition we will not acknowledge these possible actions, and consequently we will not take them. On a societal level, anthropologist Ghassan Hage (2003) writes that a lack of hope creates a society of worriers rather than carers—that is, people who are driven by fear, insecurity and fantasy rather than those driven by reality, creativity and action (see also Snow, 2018). Echoing this, recent empirical work in communication suggests that news stories that emphasize taking action about climate change make people feel more hopeful and encourage them to do something, as opposed to stories that emphasize urgency and use alarming vocabulary (e.g., *climate emergency*), which tend to make people skeptical and resistant (Feldman & Hart, 2021).

So how might we inspire hope? One might suppose that first and foremost hope requires getting all the facts. But in this paper, I will suggest that factual information is not enough. On this point, Jane Goodall offers a striking example. Known for her groundbreaking research on the social life of chimpanzees, Goodall has made it her life's work to inspire hope in younger generations on environmental issues. In a 2020 interview, Goodall remarked that people do not change their minds or become hopeful by being inundated with facts; rather such changes need to come from within, and they can be coaxed out through stories. In her words, "If you don't sit down and talk to people, how can you expect they're going to change?" (Tippett, 2020).

In this paper, we will explore this notion in more depth, with an eye toward delineating the categories of information sources that can contribute to hope and providing grounding for further research. Information sources can be defined as the entities in which information is instantiated and thus able to inform people; examples include books, articles, podcasts and other people. There is a strong tradition of research in library and information science on information sources, such as theoretical work clarifying typologies of information sources across dimensions (e.g., Kaye, 1995) and empirical work on people's preferences regarding information sources (e.g., Kim & Sin, 2011; Savolainen, 2008). Generally, this work is concerned with how people seek, find and use information to accomplish tasks, solve problems, answer questions and the like. In the present paper, however, I am examining how information can shape a person, perhaps even beneath conscious awareness, in line with prior work on the contemplative aims of information (Gorichanaz & Latham, 2019) and the interaction between information and emotion (Nahl & Bilal, 2007).

We'll begin, in the next section, by conceptualizing hope, drawing primarily on recent work done in philosophy. Following that, we will consider what philosophers and psychologists have written on the development of hope. Next, I will suggest that there are four types of information that can spark hope: information contributing to our understanding, to our imagination, to our desires, and to our metacognition. This framework provides the conceptual grounding for further research on hope and information, for example in the field of information behavior and practices, with specific suggestions for further work outlined in the conclusion.

**Defining hope: from belief to desire**

Hope is a positive attitude oriented toward a possible (yet uncertain), desired outcome. Thus hope involves a conjunction of belief and desire: a belief in what is possible, and a desire for a particular possibility (Bloeser & Stahl, 2017). Hope is not just a placid feeling; it motivates our efforts to attain the outcomes we desire, thus sometimes forming part of a self-fulfilling prophecy (Snyder, 2000). Hope may be specific, such as when we hope for particular outcomes, whether in the short term or long (e.g., hoping to get a paper accepted in a prestigious journal, hoping to go to heaven), or it may be a general disposition, what we usually describe as *hopefulness* (e.g., hoping for happiness or success for ourselves or others) (Snow, 2013).

While it is fraught to suggest that anything is a human universal, hope is at least pervasive across human societies, with a long and detailed history in numerous schools of philosophy and theology—and more recently psychology (Bloeser & Stahl, 2017; van den Heuvel, 2020). In some accounts, hope has had a negative connotation, being seen as a form of wishful thinking or naive delusion that can misdirect our efforts, or as a feeling that foolishly distracts us from the present. In other accounts, it has a positive connotation, being seen as a motivating force in the face of uncertainty. Similarly, hope has been seen as a non-rational passion by some thinkers and by others as a virtue—that is, as rational, as motivated toward morally good ends, and as reliable in helping us attain those ends (Zagzebski, 1996). In philosophical discussions of hope as a virtue, hope has been considered a general moral virtue on one hand (that is, a quality that leads us to taking morally good actions) and as an intellectual virtue on the other (that is, a quality that leads us to formulate correct beliefs and understandings) (Cobb, 2015; Łukasiewicz, 2021; Snow, 2013).

Virtues are often found at the mean between two vices; the virtue of hope has been described as sitting between the vices of presumption (what we might call *optimism*) and despair (*pessimism*) (Lamb, 2022). Indeed, though today we tend to consider optimism and hope as synonyms, this framework suggests how the two may be distinguished. As philosopher Michael Lamb (2022) contends, optimism describes a tendency to presume future goods to be certain (or to take them as certainties), whereas hope acknowledges them as possibilities, and maybe even unlikely ones, as when we "hope against hope." Moreover, true hope requires good information—about the negative possibilities as well as the positive, and about the risks associated with the possible courses of action. On this view, despair results from only being informed about the negatives, or dwelling on them, without getting suggestions for what actions can be taken or how things could be improved.

In this paper, I take the position that hope is a virtue, whether moral or intellectual, and that it is a productive response to uncertainty and difficulty. What I wish to discuss in this paper is where hope comes from in terms of information, and how information could contribute to hope as a conceptual starting point for further work in this vein across library and information studies. While discussions of the breadth of what can be information proliferate, in this paper I will focus on recorded forms of information, or information preserved in a durable medium, as defined by Bates (2006).

As a caveat, I hasten to add that hope may not always be responsive to information. In the hopelessness of certain forms of depression, for example, the underlying issue may require therapy or medical treatment. While talk therapy and medication can be considered information in an expansive sense (Day, 2021), these are outside the scope of my discussion in this paper. That said, what I say in this paper may incidentally have some relevance to talk therapy; medication, on the other hand, operates by chemical mechanisms on a different level of abstraction from what I engage with here.

**Developing hope: insights from philosophy and psychology**

Where does hope come from, and how do we develop it? This question has been taken up to some extent in both philosophy and psychology. First, philosophers of the virtues have discussed where the virtues in general come from and how we can become more virtuous; since hope is considered a virtue, presumably these strategies also apply to hope. These strategies include: being motivated to have the virtue, identifying and then emulating other people who have the virtue, and looking for opportunities to practice the virtue (Vallor, 2016; Wright et al., 2021; Zagzebski, 2017). Next, in philosophical discussions specific to hope, we can find the following sources for hope:
- reflecting on our past experiences of good fortune (Gravlee, 2000)
- reflecting on our desires (Hobbes, 1651/2021, chapter 6)
- reflecting on uncertainty or incomplete evidence (Hume, 1738/1896, §9)
- possessing other virtues, such as faith and love (Augustine, ca. 420/1955) or open-mindedness and resilience (Snow, 2013), as the virtues are mutually supportive and reinforcing
- having our basic needs met, including health, education, a strong economy, etc. (Snow, 2018)

In psychology, especially positive psychology, the question of cultivating hope has also been taken up. The recent book *Learned hopefulness* (Tomasulo, 2020) offers the most comprehensive statement available on the subject. According to Tomasulo, strategies for cultivating hope include: understanding the nature of hopelessness, learning to see possibilities, cultivating positive feelings and observations, setting meaningful goals, acting as if we are already hopeful, and nurturing personal relationships. This advice is echoed in professional clinical literature (Lopez et al., 2004), as well as general-audience articles from other psychologists (e.g., Casablanca, 2021; Jed Foundation, n.d.; Leahy, 2021; Morin, 2021; Zapata, 2019),.

Considering all these strategies together, the focus has been on what we can achieve through individual cognition, with a few exceptions. If we are interested in understanding how exosomatic information could be used to spark or foster hope, these authors have said little, though there are implicit references to exosomatic information in some of their recommendations, such as in recommendations to '*understand what hopelessness is*' (Zapata, 2019) and '*review the evidence that positive changes… are possible*' (Leahy, 2021), as well as in terms of reflecting on evidence.

**Theorizing information sources for hope**

Let's take a step back from the particular the suggestions above to theorize sources for hope. This theorization takes the form of conceptual analysis, a philosophical method of breaking down a concept or theory into its constituent parts for further examination. For a primer on conceptual analysis as applied to questions of library and information studies, see Furner (2004).

Recall that hope, at heart, involves two components: belief in what is possible and desire for a particular possibility. There is also a third piece involved, which has perhaps gone without saying: belief about the present. Any hope for the future must be grounded in a particular present out of which that hope could spring. Thus hope involves a relationship between what is, what could be, and what we want. Hope can be facilitated, then, by introducing information at any of these three places:
1. **Understanding**: information for creating beliefs about the present or past
2. **Moral imagination**: information for opening up the imagination about future possibilities
3. **Desire**: information creating desires among future possibilities as part of pursuing a life of flourishing

I would suggest that, to have hope, we must be well-informed in all three of the above dimensions. To further expand our hope, we could get information that feeds back into other information, such as introducing ambiguity or revealing the mechanisms of hope; we might describe this as meta-information, or information that informs our metacognition (how we think about our thinking):
4. **Metacognition**: information about our information, for revealing the incompleteness of our evidence, instilling a sense of ambiguity, demonstrating how hope works, etc.

*Information for understanding*

The first category of information for hope is that which affects our beliefs about the world, whether present or past (insomuch as the past bears upon the present). To be conducive for hope, such information needs to provide solid ground for hope to stand upon. That is, it must imply or demonstrate that purely negative outcomes in the future are not already a foregone conclusion.

To give an example regarding climate change, a book promising that possibilities for life will be utterly annihilated in a matter of decades and there's nothing we can do about it will not in any way inspire hope (or, importantly, the constructive action that hope motivates). A recent popular book close to this description is *The uninhabitable earth: life after warming*, by David Wallace-Wells (2019), which paints a number of apocalyptic scenarios for the near future, ranging from bad to worse. But if a book demonstrates that our dire predictions are not yet certainties, that there is still room for decisive human action, then hope has ground to stand upon. A recent book along these lines is Hans Rosling's (2018) *Factfulness: ten reasons we're wrong about the world—and why things are better than you think,* which shares statistics and principles about why we think our current circumstances, such as regarding the climate, are bad and seeks to disabuse us of false beliefs and biases (e.g., us vs. them tendencies, fear-driven media). To be sure, hope inspired by such information is only true hope (rather than false hope) if it is rooted in genuine information—contextualized and complete, rather than misleading strands of statistical sophistry.

This type of information could be described as factual information, i.e., what philosopher Luciano Floridi describes as strongly semantic information (Floridi, 2011). But we can easily imagine (or remember from our own experience) situations where a person has all the relevant facts about some matter but they remain hopeless. In such cases, to become hopeful, we need different kinds of cognitive resources beyond factual information about the present or past. In addition to facts, hope seems to require what we might broadly describe as story—that is, the way those facts fit together in a narrative scheme. This leads us to the next category of information for hope.

*Information for moral imagination*

Author Jacqueline Novogratz defines moral imagination as '*the humility to see the world as it is, and the audacity to imagine the world as it could be*' (Novogratz, 2020, p. 239). Our moral imagination is our horizon for the good that may be possible. Just as with eyesight, some of us morally see farther and more sharply than others. That is, some of us have narrow moral imaginations, while others' moral imaginations are more expansive. Our moral imagination is not an inborn, fixed trait, but rather it can change and grow (Johnson, 1993).

Information can influence our moral imagination, but it is not (just) strongly semantic information that does so. Rather, we need what Floridi (2018) has called *semantic capital*. Semantic capital is any content that can enhance a person's power to give meaning to something, a process Floridi terms *semanticization*. Put differently, semantic capital is the kind of cognitive resource that we need to answer open questions, i.e., questions of value, preference, etc. (Floridi, 2019). Such resources include one's cultural background, religion, art, social practices, expectations, emotions and so on.

While I would argue that our semantic capital bears on how we understand *anything*, in this paper my scope is limited to making the point that our semantic capital helps define our moral imagination. An example of this is given by philosopher Jonathan Lear in *Radical hope:*

*ethics in the face of cultural devastation* (Lear, 2006). In this book, Lear observes that our understandings of ourselves and our possibilities occur within the rubric of our particular ways of life (cultural vocabulary, rituals, etc.), and he asks how we might be able to move forward when our way of life falls apart. His answer is *radical hope*, a flexibility that is tuned to reality but not tied down by it. For Lear, such hope is built through *poetry*, which he defines broadly as the creative making of meaningful space (Lear, 2006, p. 51).

Lear develops these ideas through the case of Plenty Coups (1848–1932), an American Indian Crow chief, who led the Crow people as they transitioned from their traditional way of life to a new one inflected by contemporary American culture. On Lear's account, the Crow were able to make this transition because of how they engaged the practice of vision quests. He writes, '*The Crow had an established practice for pushing at the limits of their understanding: they encouraged the younger members of the tribe (typically boys) to go off into nature and dream*' (Lear, 2006, p. 66). The young people returned to the tribe and shared their dreams, which were then interpreted within the tribal context as part of their decision-making in effort to bridge past traditions with present circumstances. Lear credits Plenty Coups' ability to do this as a reason for the Crow's continued existence; he contrasts Plenty Coups with Sioux chief Sitting Bull (1831–1890), who held too fast to the letter of past traditions and avoided the real-life demands of a changing world (Lear, 2006, p. 151).

To put this another way, the Crow people used dreams and visions, intentionally retrieved from the fuzzy borderlands of the horizon of possibilities, as ingredients in broadening their moral imagination. This involves transformation, but with continuity. '*For if one is straining to live, and help others, in a worthwhile way, the question can no longer be, say, "How shall I, as a Crow, go on?" but "What shall it be for me to go on as a Crow?" And one is forced to address this question at a time when it is no longer clear how one could possibly answer it*' (Lear, 2006, pp. 47–48).

The information in this category is that which broadens our horizons, inviting us to see possibilities that hadn't occurred to us. The broader one's moral imagination, the more expansive and developed it is. Continuing with the geographic metaphor of "horizon," the moral imagination constitutes a swathe of houses, which can then be furnished with facts, the sort of information discussed in the previous section. And just as roads and maps are tools that help us navigate our real-life landscapes, so we need tools to navigate the horizon of our moral imagination. This leads us to the next section.

### *Information for desire*

Hope requires not just belief and imagination, but also desire. We cannot hope for something we don't want. Information can certainly expand our desires. An obvious example in the commercial realm is advertising, through which we learn about new products and come to want them. In the words of writer David Foster Wallace, advertisements '*create an anxiety relievable by purchase*' (Wallace, 1998, p. 414). We may use the term *hope* to describe desires in the commercial realm—as in hoping for a Christmas present or to be able to afford a new purchase—but when it comes to hope as a virtue, the desire must be for something deeper than just the latest iPhone.

Desires that are conducive to hope are those related to *the good life*. And though the good life certainly involves a level of material accrual—we might recall the proverbial phrase '*healthy, wealthy and wise*'—having possessions is not the only thing. That said, beyond the common observation that material accrual is not enough, what exactly constitutes the good life is an open question (read: it requires semantic capital to answer, and it cannot be decisively answered). To give one statement on the matter, MacIntyre (1981/2007) in his groundbreaking study of moral theory defines the good life in this way:

> *The good life for man is the life spent in seeking for the good life for man, and the virtues necessary for the seeking are those which will enable us to understand what more and what else the good life for man is.* (MacIntyre, 2007, p. 219)

To put this in more cliche terms, the good life may be found more in the journey than the destination.

What sort of information can provide inspiration and stoke desire for the seeking that is the good life? According to virtue theorists, the major information source for the good life is other people, particularly those people who are already living well. Such people are called *moral exemplars*: the people who we admire upon reflection and seek to emulate in our own attitudes and behaviors (Zagzebski, 2017). The phrase '*upon reflection*' is vital. For example, we may at first blush admire the wealth of Jeff Bezos or Elon Musk, but after deep consideration we may realize we do not wish to emulate these men (though here I leave that question aside).

Emulating moral exemplars is the explicit structure of certain religions, such as Christianity and Buddhism. In Christianity, the figure of Christ is the ultimate moral exemplar; his followers emulated him, and their followers have emulated them. '*Be imitators of me, just as I am also of Christ',* writes Paul in 1 Corinthians 11. Particularly in the Roman Catholic tradition of Christianity, As are held in particular esteem as moral exemplars. MacIntyre (2007, p. 199) suggests several saints that may serve in this regard, even to secular persons: Benedict, Theresa and Francis. In Buddhism, there is a similar structure in the transmission of the Dharma through a lineage of teachers, tracing back to the Buddha himself; just as with Christianity, Buddhist traditions also recognizes saints such as Mahakasyapa and Upagupta (Ray, 1994). But moral exemplarist theory is not bound to any particular religious tradition, and secular persons may serve as moral exemplars. Likewise, as MacIntyre also suggests, certain religious figures may be morally admired by people of other faiths or of no faith. For example, Gandhi may serve as a moral exemplar even to non-Hindus (Zagzebski, 2017).

If moral desire can be awakened through identifying and observing moral exemplars, then sources that provide information in this regard include stories of good lives: certain religious texts, biographies and memoirs, as well as certain fictional narratives. To give one example, we might consider a biography of Saint Francis of Assisi (ca. 1181–1226), such as the classic written by G. K. Chesterton (1923), in this regard. A charismatic leader, Francis led the movement toward evangelical poverty and humility, and he is now the patron saint of animals and the natural environment.

### *Information for metacognition*

In addition to the above three types of information sparking belief, imagination and desire, hope can be fostered by a fourth category of information, which is information aimed at a person's metacognition, or our ability to reason about reasoning (Catalano, 2017). Such information in this case may include academic research from philosophy, psychology and theology, *inter alia*, explaining the mechanisms of hope, providing reasons to embrace uncertainty, and so on. In this category fit the works sketched above in the section "Developing hope," as well as many others.

To cite just one example here, we might consider William James' (1912/2009) essay "The will to believe," which discusses the extent to which we can choose our beliefs, the limits of our possibilities for certainty in life, the relationship between faith and fact, and the circumstances in which it may be wise to choose to believe in unpopular or unlikely possibilities.

**Conclusion**

This paper has provided a theoretical account of the information sources for hope, which fall into four categories: information inspiring understanding, imagination, desire, and metacognition. If hope is indeed a virtue, as any number of scholars both present and past have argued, then we in library and information studies should help connect people with the information that will inspire hope.

This paper is a step in that direction. Future work can provide a fuller account of information and hope, particularly as informed hope fuels and directs further searching for information and eventual action. Such work may form part of a broader empirical research program on the interactions between virtue and people's information behavior and practices. How do hopeful people seek and experience information? What is the fine-grained texture of hopeful information seeking? How does hope connect with other virtues when it comes to information needs, experience and creation? To develop studies to answer such questions, classic research methods from information behavior and practices could be harnessed; for example, a researcher could ask people to reflect on moments when they felt hopeful or hopeless and explore their experiences in a survey or interview to identify where and how information was involved in those experiences.

To close, I want to sketch an example from my own life, one that is still unfolding, as a sort of nascent case study. Lately I have been disturbed by the rising gun violence where I live, in a major American city (more details to be provided after blind review). This hasn't just been an intellectual matter. Earlier this year, there was a shooting on my block while I was outside, and I found myself holding one of the victims steady while we waited for the police and ambulance to arrive. This experience led me to seek out news and statistics about gun violence and homicide in 2020 and 2021, particularly in my city. Meanwhile in the following months there were two more shootings, and I was not hopeful that anything would get better. I found myself continually on-edge while walking outside, jumping at sudden noises. Perhaps instinctively, I have been looking for reasons to be hopeful. I've found myself drawn to learning more about urban gun violence, including its causes and proposals for dealing with it. There seems to be a fine line between getting the information needed to form accurate beliefs (the foundation of hope), and getting so much information that I slip into despair. Along the way, I've learned about initiatives coming from the city, state and national governments, as well as about community organizations and efforts going on in other cities. Only very recently has any hope begun to spring up within me. I've learned that there are many, many people concerned about this issue, and efforts on many levels working to address it. I have found myself called to contribute in the ways that I can, which for me center on community-based mentorship programs for vulnerable youth. All this is to say that my own experience has provided a case study for the relationship between information, hope and action, a thread that I hope others will pick up on.

**About the author**

Tim Gorichanaz is an Assistant Teaching Professor in the College of Computing & Informatics at Drexel University, where he teaches courses across information science, human–computer interaction, and design. He received his PhD from Drexel University, and his research interests are in information experience, information ethics and document theory. He can be contacted at gorichanaz@drexel.edu.